\documentclass[aps,preprint,showpacs,groupedaddress,floatfix]{revtex4}

\usepackage{amssymb}
\usepackage{graphicx}
\usepackage{dcolumn}
\usepackage{bm}
\usepackage{CJK}
\usepackage{float}

\begin{document}
\begin{CJK*}{GBK}{song}

\title{Reexamining the temperature and neutron density conditions for $r$-process
nucleosynthesis with augmented nuclear mass models}
\author{X. D. Xu$^1$}
\author{B. Sun$^{1,2}$}\thanks{e-mail: bhsun@buaa.edu.cn}
\author{Z. M. Niu$^3$}
\author{Z. Li$^1$}
\author{Y.-Z. Qian$^{4,1}$}\thanks{e-mail: qian@physics.umn.edu}
\author{J. Meng$^{1,5,6}$}\thanks{e-mail: mengj@pku.edu.cn}
 \affiliation{$^1$School of Physics and Nuclear Energy Engineering, Beihang University, Beijing 100191, China}
 \affiliation{$^2$Justus-Liebig-Universit\"{a}t Giessen, Heinrich-Buff-Ring 14, Giessen 35392, Germany}
 \affiliation{$^3$Department of Physics, Anhui University, Hefei 230601, China}
 \affiliation{$^4$School of Physics and Astronomy, University of Minnesota, Minneapolis, Minnesota 55455, USA}
 \affiliation{$^5$State Key Laboratory of Nuclear Physics and Technology, School of Physics, Peking University,
 Beijing 100871, China}
 \affiliation{$^6$Department of Physics, University of Stellenbosch, Stellenbosch, South Africa}

\date{\today}
\begin{abstract}
We explore the effects of nuclear masses on the temperature and neutron density
conditions required for $r$-process nucleosynthesis using four nuclear mass models
augmented by the latest atomic mass evaluation. For each model we derive the conditions for producing
the observed abundance peaks at mass numbers $A\sim 80$, 130, and 195 under the waiting-point approximation
and further determine the sets of conditions that can best reproduce the $r$-process abundance patterns
($r$-patterns) inferred for the solar system and observed in metal-poor stars of the Milky Way halo.
In broad agreement with previous studies, we find that (1) the conditions for producing
abundance peaks at $A\sim 80$ and 195 tend to be very different, which suggests that, at least
for some nuclear mass models, these two peaks are not produced simultaneously;
(2) the typical conditions required by the critical waiting-point (CWP) nuclei with the $N = 126$
closed neutron shell overlap significantly with those required by the $N=82$ CWP nuclei, which
enables coproduction of abundance peaks at $A\sim 130$ and 195 in accordance with
observations of many metal-poor stars; and (3) the typical conditions required
by the $N = 82$ CWP nuclei can reproduce the $r$-pattern observed in the metal-poor star
HD~122563, which differs greatly from the solar $r$-pattern. We also examine how nuclear
mass uncertainties affect the conditions required for the $r$-process and identify
some key nuclei including $^{76}$Ni to $^{78}$Ni, $^{82}$Zn, $^{131}$Cd, and
$^{132}$Cd for precise mass measurements at rare-isotope beam facilities.
\end{abstract}
\maketitle
\date{today}
%%%%%%%%%%%%%%%%%%%%%%%%%%%%%%%%%%%%%%%

\section{Introduction}\label{sec:introduction}
Nucleosynthesis via rapid neutron capture, the $r$-process, is a
major mechanism for producing the elements heavier than
Fe~\cite{Burbidge1957,Cameron1957}. Understanding this process
requires knowledge of properties such as masses, $\beta$-decay
lifetimes, and neutron-capture cross sections for a large number of
extremely neutron-rich nuclei far from stability (e.g.,
~\cite{Cowan1991,Qian2003,Arnould2007}). Most of this nuclear input
is beyond the reach of experiments in the foreseeable future and,
therefore, must be calculated with guidance from existing data and
from measurements to be made at rare-isotope beam facilities such as
the GSI Facility for Antiproton and Ion Research (FAIR), the
Facility for Rare Isotope Beams (FRIB), the Heavy Ion Research Facility
in Lanzhou Cooling Storage Ring (HIRFL-CSR), and Rikagaku Kenkyusho
(RIKEN, Institute of Physical and Chemical Research, Japan).
In this paper we explore the importance of nuclear masses in estimating
the temperature and neutron density conditions required for a specific $r$-process
scenario, where neutron-capture reactions are in equilibrium with
the reverse photodisintegration reactions; i.e., there is
$(n,\gamma)\rightleftharpoons(\gamma,n)$ equilibrium
(e.g.,~\cite{Cowan1991,Qian2003,Arnould2007}). Using four nuclear
mass models, we show that the required conditions can be determined
mostly from the neutron-separation energies for a small number of
critical nuclei with $N=50$, 82, and 126 closed neutron shells. For
each model, we further determine the best-fit sets of conditions to
reproduce the $r$-process abundance pattern ($r$-pattern) inferred
for the solar system and observed in metal-poor stars of the Milky
Way halo. This allows us to draw several interesting conclusions
regarding the production of different parts of the overall
$r$-pattern in an $(n,\gamma)\rightleftharpoons(\gamma,n)$ equilibrium scenario. We
also illustrate the effects of nuclear mass uncertainties on the
required $r$-process conditions and identify the key nuclei that
have the largest impact and, therefore, are important candidates for
precise mass measurements at rare-isotope beam facilities.

We first give a brief overview of the $r$-process. Detailed reviews can be found
in Ref.~\cite{Cowan1991,Qian2003,Arnould2007}. Historically, the abundance distribution
of nuclei in the solar system played a crucial role in studies on the origin of the
elements~\cite{Burbidge1957,Cameron1957}. One of the prominent features of this distribution is the
presence of three sets of double peaks in the region beyond the Fe group nuclei.
This was recognized as signatures of two distinct processes of neutron capture: a slow ($s$)
one encountering the $N=50$, 82, and 126 closed neutron shells in the stable region and
a rapid ($r$) one encountering the same in the extremely neutron-rich region of the nuclear
chart~\cite{Burbidge1957,Cameron1957}. Specifically, the peaks at mass numbers $A\sim 80$,
130, and 195 were produced by the $r$-process and represent the crucial features of the
solar $r$-pattern, which is derived by subtracting the $s$-process contributions from
the net solar abundances (e.g.,~\cite{Kappeler2011}).

In order to fully understand the $r$-process, we need conditions such as temperature and
neutron density in the associated astrophysical
environments in addition to the properties of a large number of extremely neutron-rich nuclei.
Neither the astrophysical nor the nuclear input is firmly established, although much progress
has been made over the past two decades~\cite{Cowan1991,Qian2003,Arnould2007}.
Proposed astrophysical sites for the $r$-process include neutrino-driven winds from
proto-neutron stars formed in core-collapse supernovae (CCSNe)~\cite{Woosley1992,Meyer1992,Woosley1994},
shocked surface layers of O-Ne-Mg cores associated with low-mass CCSNe~\cite{Ning2007},
winds from accretion disks of black holes formed in high-mass CCSNe~\cite{Pruet2003,Surman2006,Surman2008,Wanajo2012},
He shells of metal-poor CCSNe~\cite{Epstein1988,Banerjee2011}, and
ejecta from neutron star mergers~\cite{Lattimer1977,Freiburghaus1999a,Goriely2011,Korobkin2012}.
There are large uncertainties in the conditions associated with all CCSNe environments
due to the substantial uncertainties in modeling such environments (e.g.,~\cite{Janka2012}),
especially when neutrino transport in hot and dense nuclear matter is considered
(e.g.,~\cite{Martinez2012,Roberts2012}). While recent studies lend much support to neutron star mergers
being an $r$-process site~\cite{Goriely2011,Korobkin2012}, it remains to
be seen whether such models are consistent with the history of $r$-process enrichments in the Milky Way
and in its satellite dwarf galaxies (e.g.,~\cite{Qian2000,Argast2004,Donder2004}). Further,
how sensitive these models are to the uncertainties in the current understanding of the
nuclear equation of state (e.g.,~\cite{Page2006}) and to the numerical treatment of
the merger dynamics remains to be studied in detail.

The conditions in the astrophysical environments relevant for the $r$-process ultimately
boil down to the seed nuclei for neutron capture at the beginning of the process
and the temperature $T(t)$ and the neutron (number) density $n_n(t)$ as functions of time
$t$ during the process (e.g.,~\cite{Freiburghaus1999b}).
In cases where neutrino interactions are important, the time evolution of neutrino fluxes and
energy spectra is also required (e.g.,~\cite{Epstein1988,Banerjee2011}). In the rest of the paper
we will ignore neutrinos and focus on a broad class of astrophysical environments where matter
undergoes $r$-processing at $T\gtrsim 10^9$~K and $n_n\gtrsim 10^{20}$~cm$^{-3}$.
For such high temperatures and neutron densities, previous
studies have shown that $(n,\gamma)\rightleftharpoons(\gamma,n)$ equilibrium is
achieved (e.g.,~\cite{Goriely1996}). In this equilibrium, the abundance distribution in each
isotopic chain at a specific proton number $Z$ is almost always strongly peaked at one nucleus.
This is referred to as a waiting-point (WP) nucleus because, upon reaching it, the $r$-process
must wait for it to $\beta$-decay before producing heavier nuclei. Under this so-called
WP approximation, the $r$-process path is defined by all the WP nuclei heavier than
the seed nuclei, and the progress along this path is regulated by the $\beta$-decay
of these WP nuclei. So long as this approximation is valid, there is no need to follow
neutron capture and photodisintegration reactions, which greatly simplifies the
$r$-process calculation.

Due to the equilibrium between neutron capture and photodisintegration reactions,
the abundance ratio between two neighboring isotopes is given by the Saha equation
(e.g.,~\cite{Cowan1991,Qian2003,Arnould2007}):
\begin{equation}\label{equilibrium}
\frac{Y(Z,A+1)}{Y(Z,A)}=n_n\left(\frac{2\pi\hbar^2}{m_u k T}\right)^{\frac{3}{2}}
\frac{G(Z,A+1)}{2G(Z,A)}\left(\frac{A+1}{A}\right)^{\frac{3}{2}}
\exp\left[\frac{S_n(Z, A+1)}{ k T}\right],
\end{equation}
where $\hbar$ is the Planck constant, $m_u$ is the atomic mass unit, $k$ is the Boltzmann
constant, $(Z,A)$ indicates a nucleus with proton number $Z$ and mass number $A$,
and $Y$, $G$, and $S_n$ denote the number abundance, partition function, and neutron
separation energy of the appropriate nucleus, respectively.
For a specific isotopic chain, the corresponding WP nucleus has the largest abundance and
is determined by the partition functions and neutron separation energies of the relevant
nuclei for fixed $T$ and $n_n$. As can be seen from the exponential dependence on
the neutron separation energy in Eq.~(\ref{equilibrium}), nuclear masses are among the most
important input for modeling the $r$-process. The other crucial input is $\beta$-decay lifetimes
of the relevant nuclei.

Over the past two decades, tremendous progress has been made in measuring nuclear
properties relevant for the $r$-process. For example, the $\beta$-decay half-lives of 38
very neutron-rich isotopes bordering the $r$-process path have been measured
recently~\cite{Nishimura2011}. In addition, the masses of a group of nuclei including
$^{80}$Zn~\cite{Baruah2008, Sun2008b} and $^{130}$Cd~\cite{Dillmann2003} have been
measured with a very high accuracy~\cite{Audi2011}. Meanwhile, considerable advance
has been made in the theoretical investigation of nuclear masses. The four nuclear mass
models used in this paper span from the macroscopic-microscopic kind, represented by
the finite-range droplet model (FRDM)~\cite{Moller1995} and a more recent
Weizs\"acker-Skyrme (WS*) model~\cite{Wang2010},
to the microscopic kind, represented by the Skyrme-Hartree-Fock-Bogolyubov mean-field
(HFB-17) model~\cite{Goriely2009} and the relativistic mean-field (RMF) model~\cite{Geng2005}.
These models can reproduce the experimentally known neutron separation energies
with a root-mean-square (rms) deviation of 0.399 (FRDM), 0.332 (WS*), 0.506 (HFB-17), and
0.653 (RMF) MeV, respectively.

Based on the above overview, there are two frontiers of $r$-process research: one focusing
on the search for the astrophysical sites and quantification of the conditions therein and
the other on acquiring a reliable database for the relevant nuclear input. Observations of
elemental abundances in metal-poor stars of the Milky Way halo (see~\cite{Sneden2008}
for a review) have shed important light on the $r$-process sites (e.g.,~\cite{Qian2007,Farouqi2009}).
The observed $r$-patterns also provide an important test of the basic soundness of the nuclear
input (e.g.,~\cite{Farouqi2010}). Of course, the astrophysical and the nuclear input must be
coupled together in order to produce an $r$-pattern for comparison with observations.
With substantial uncertainties in the current understanding of both the $r$-process sites
and the nuclear input, parametrization of the astrophysical conditions is often used in
exploring the effects of nuclear input on $r$-process production
(e.g.,~\cite{Freiburghaus1999b,Farouqi2010}). As a practical matter, we adopt the classical
approach of using $T$, $n_n$, and the corresponding neutron irradiation time $\tau$ along
with the WP approximation (e.g.,~\cite{Kratz1993}) to carry out our $r$-process calculations below.
Our main purpose is to explore the effects of the four nuclear mass models mentioned above
on the $T$ and $n_n$ conditions required for $r$-process nucleosynthesis.

A number of other studies on how the nuclear input impacts $r$-process nucleosynthesis
have been carried out recently. The influence of nuclear properties from different mass models
on the final $r$-pattern was analyzed in Ref.~\cite{Farouqi2010}. The effect of neutron capture
rates for nuclei near the $A\sim 130$ peak on the overall $r$-pattern was investigated in Ref.~\cite{Surman2009}.
The sensitivity of the calculated $r$-pattern to the combined effects of the long-term dynamic evolution
of the astrophysical environment and the nuclear input was explored in Ref.~\cite{Arcones2011a}.
The effect of long-range correlations for nuclear masses on the production of nuclei immediately
below and in the $A\sim 195$ peak was studied in Ref.~\cite{Arcones2012}.
In all of the above studies parametric astrophysical models that allow $T$ and $n_n$ to evolve with time were adopted.
These models are more realistic but the choice of parameters is not so straightforward.
In the future, we plan to use similar models to study the interplay between the astrophysical
and the nuclear input during the $r$-process. Our goal here is to explore the $T$ and $n_n$ conditions
required for the $r$-process in an $(n,\gamma)\rightleftharpoons(\gamma,n)$ equilibrium scenario and
the effects of nuclear mass models on these conditions. For this limited goal, we feel that the
classical approach based on the WP approximation is adequate.

We give a detailed discussion of the WP approximation in Sec.~\ref{sec:model}.
Using this approximation along with four nuclear mass models, we derive for each model
the $T$ and $n_n$ conditions that are required for producing the abundance peaks at
$A\sim 80$, 130, and 195 as observed in the solar system.
In Sec.~\ref{sec:patterns} we describe the classical approach to simulate the $r$-process
and use this approach to determine the sets of conditions that can best reproduce
the solar $r$-pattern and the $r$-patterns observed in metal-poor stars for each
of the adopted nuclear mass models. We discuss our results and give conclusions in
Sec.~\ref{sec:discuss}.

%--------------------------------------------------------------------------------------------
\section{$r$-Process Conditions under the WP Approximation}\label{sec:model}
%--------------------------------------------------------------------------------------------
While there are substantial uncertainties in both the $r$-process sites and the relevant
nuclear input, an essential feature of the $r$-process is considered robust: the observed
abundance peaks at $A\sim 80$, 130, and 195 correspond to the intrinsic properties of
extremely neutron-rich nuclei with $N=50$, 82, and 126 closed neutron shells that are
produced in the $r$-process. As discussed in Sec.~\ref{sec:introduction}, when
$(n,\gamma)\rightleftharpoons(\gamma,n)$ equilibrium is achieved, the total
abundance of an isotopic chain is concentrated in the corresponding WP nucleus.
The $\beta$-decay lifetimes of the WP nuclei then regulate the abundance pattern
resulting from an $r$-process episode. In particular, the much longer $\beta$-decay
lifetimes of extremely neutron-rich nuclei with closed neutron shells than those without
produce peaks in $r$-patterns. Consequently, in order to produce the observed peaks
in $r$-patterns under the WP approximation, nuclei with $N=50$, 82, and 126
closed neutron shells and with $A\sim 80$, 130, and 195, respectively, must be among
the WP nuclei. The critical WP (CWP) nuclei were discussed by earlier studies
(e.g.,~\cite{Kratz1988}). For the present work, we select the CWP nuclei listed in
Table~\ref{tabCWP} based on similar considerations to these studies.
Below we follow the spirit of previous studies to derive the conditions required for $r$-process
nucleosynthesis under the WP approximation by considering the properties of the CWP nuclei.

%-------------------------------------------------------------------------------------
\begin{table}[h]
  \renewcommand{\arraystretch}{1.5}
  \centering
  \caption{Critical waiting-point nuclei.}\label{tabCWP}
  \setlength{\tabcolsep}{20pt}
  \vspace{1em}
    \begin{tabular}{ll}
      \hline
      \hline
      $N$ & CWP nuclei\\
      \hline
      50 & $^{80}$Zn, $^{79}$Cu, $^{78}$Ni\\
      82 & $^{130}$Cd, $^{129}$Ag, $^{128}$Pd, $^{127}$Rh, $^{126}$Ru\\
      126 & $^{195}$Tm, $^{194}$Er, $^{193}$Ho, $^{192}$Dy, $^{191}$Tb\\
      \hline
      \hline
    \end{tabular}
\end{table}
%---------------------------------------------------------------------------------------

%---------------------------------------------------------------------------------------
\subsection{Role of neutron separation energies}\label{sec:nsep}
%--------------------------------------------------------------------------------------
To be quantitative, we define a WP nucleus $(Z,A_{\rm WP})$ as one that has an
abundance
\begin{equation}
Y(Z,A_{\rm WP})\geqslant 0.5\sum_AY(Z,A),
\end{equation}
where the sum over $A$ gives the total abundance of the
corresponding isotopic chain. For specific $T$ and $n_n$, we can use the above criterion
and the relative abundance $Y(Z,A+1)/Y(Z,A)$ given by Eq.~(\ref{equilibrium}) to
determine $(Z,A_{\rm WP})$ from the nuclear partition functions and neutron separation
energies provided by a model. Conversely, we can also determine the $T$ and $n_n$
conditions required by a specific WP nucleus. As can be seen from Eq.~(\ref{equilibrium}),
the predominant dependence of $Y(Z,A+1)/Y(Z,A)$ is on the neutron separation energy,
which can be calculated from a nuclear mass model. We ignore the small differences in
the nuclear partition function and in the mass number and rewrite Eq.~(\ref{equilibrium}) as
\begin{equation}\label{equilibrium2}
\frac{Y(Z,A+1)}{Y(Z,A)}=\exp\left[\frac{S_n(Z, A+1)-S_n^0(T,n_n)}{k T}\right],
\end{equation}
where
\begin{equation}\label{Sn0}
S_{n}^{0}(T,n_n)\equiv kT\ln\left[\frac{2}{n_n}\left(\frac{m_u kT}{2\pi\hbar^2}\right)^{3/2}\right]
=T_9\left[2.79+\frac{1.5\log_{10} T_9-\log_{10}(n_n/10^{20}\ {\rm cm}^{-3})}{5.04}\right]\,\rm{MeV}.
\end{equation}
In the second equality of Eq.~(\ref{Sn0}), $T_9$ is $T$ in units of $10^9$~K.
Equation~(\ref{equilibrium2}) is used in the calculations below.

\begin{figure}[h]
\centerline{
\includegraphics[scale=0.28, angle=0]{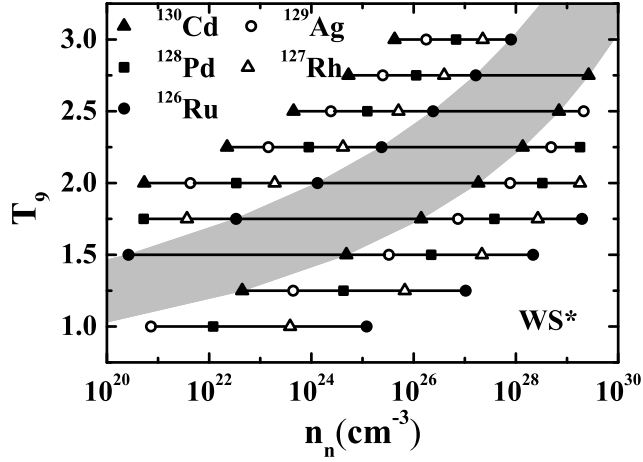}
}
\caption{The $T_9$-$n_n$ conditions required by the $N = 82$ CWP nuclei
based on the WS* mass model. For a specific $T_9$, the values of $n_n$ between
two identical symbols would allow the corresponding nucleus to have $\geqslant50\%$
of the total abundance of its isotopic chain. The conditions indicated by the shaded
band are required to accommodate all the $N = 82$ CWP nuclei.}
\label{figWSpure82}
\end{figure}

As an example, we adopt the WS* mass model to calculate the sets of $T_9$ and $n_n$
within the ranges $1 \leqslant T_9 \leqslant 3$ and $10^{20} \leqslant n_n \leqslant 10^{30}$~cm$^{-3}$
that are required by the $N=82$ CWP nuclei. The results are shown in Fig.~\ref{figWSpure82}.
For a specific $T_9$, the values of $n_n$ between two identical symbols in this figure
would allow the corresponding nucleus to have $\geqslant50\%$ of the total abundance of
its isotopic chain. In order to accommodate all the $N=82$ CWP nuclei, the common
range of $n_n$ for a specific $T_9$ is bounded from below by $^{126}$Ru (filled circle)
and from above by $^{130}$Cd (filled triangle). This range of $n_n$ changes with $T_9$
and is shown as the shaded band in Fig.~\ref{figWSpure82}. This band represents
the $T_9$-$n_n$ conditions required by the $N=82$ CWP nuclei.

%--------------------------------------------------------------------------------------
%--------------------------------------------------------------------------------------

\begin{figure}[h]
\centerline{
\includegraphics[scale=0.28,angle=0]{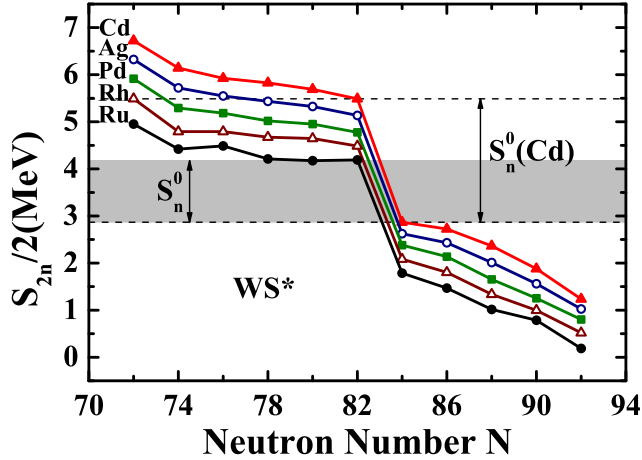}
}
 \caption{(Color online) Two-neutron separation energies $S_{2n}$ for
even-$N$ isotopes of Ru, Rh, Pd, Ag, and Cd ($44\leqslant Z\leqslant 48$) in the WS* model
shown in terms of $S_{2n}/2$ as a function of $N$ around $N = 82$.
The band between the dashed lines labeled as $S_{n}^{0}({\rm Cd})$ corresponds
to the $T_9$-$n_n$ conditions required for $^{130}$Cd to be a WP nucleus.
Likewise, the shaded band labeled as $S_{n}^{0}$ corresponds to the
conditions required to accommodate all the $N=82$ CWP nuclei
(shaded band in Fig.~\ref{figWSpure82}). See text for details.}
\label{figS2n}
\end{figure}

While the neutron separation energies of a large number of nuclei from
the WS* mass model are used in the above calculations, the results in
Fig.~\ref{figWSpure82} are determined effectively by the two-neutron
separation energies of the $N=82$ CWP nuclei and their $N=84$
isotopes due to nuclear systematics. This can be understood as follows.
Because of the effect of pairing on neutron binding, all WP nuclei have
even $N$. The relative abundance of two neighboring even-$N$
isotopes can be obtained from Eq.~(\ref{equilibrium2}) as
\begin{equation}\label{equibibriumEven}
\frac{Y(Z, A)}{Y(Z, A-2)}=\exp\left[\frac{S_{2n}(Z, A)-2S_{n}^{0}(T,n_n)}{kT}\right],
\end{equation}
where $S_{2n}$ denotes the two-neutron separation energy.
The values of $S_{2n}/2$ for isotopes of Ru, Rh, Pd, Ag, and Cd ($44\leqslant Z \leqslant48$)
are shown as functions of $N$ in Fig.~\ref{figS2n}, which exhibits the general trend
that $S_{2n}$ essentially monotonically decreases with $N$ for a specific isotopic chain.
Based on this aspect of nuclear systematics, it can be seen from
Eq.~(\ref{equibibriumEven}) that the abundance of an even-$N$ isotope increases with $N$
[$Y(Z, A)/Y(Z, A-2)>1$] until $S_{2n}(Z,A)/2$ falls below $S_{n}^{0}(T,n_n)$, from which
point on it decreases with $N$ [$Y(Z, A)/Y(Z, A-2)<1$]. Therefore, the abundance of
even-$N$ isotopes peaks at the nucleus $(Z,A_{\rm WP})$, for which
\begin{equation}
S_{2n}(Z, A_{\rm WP}+2)\leqslant 2S_{n}^{0}(T,n_n)\leqslant S_{2n}(Z, A_{\rm WP}).
\label{S2nWP}
\end{equation}
The above equation effectively defines a WP nucleus (e.g.,~\cite{Goriely1992})
and can be used to determine the $T_9$-$n_n$ conditions required by a specific
WP nucleus. For example, $^{130}$Cd has $S_{2n}/2=5.488$~MeV while
$^{132}$Cd has $S_{2n}/2=2.869$~MeV~\cite{Wang2010}.
So the $T_9$-$n_n$ conditions corresponding to
$2.869\leqslant S_{n}^{0}(T,n_n)\leqslant 5.488$~MeV [the band between dashed lines labeled
as $S_n^0({\rm Cd})$ in Fig.~\ref{figS2n}] are required for $^{130}$Cd to be
a WP nucleus. Likewise, the shaded band labeled as $S_n^0$ in Fig.~\ref{figS2n}
corresponds to the conditions required to accommodate all the $N=82$ CWP
nuclei (shaded band in Fig.~\ref{figWSpure82}).

%--------------------------------------------------------------------------
%----------------------------------------------------------------------------

%---------------------------------------------------------------------------------------
\subsection{$T_9$-$n_n$ conditions for four nuclear mass models}\label{sec:cond}
\begin{figure}[h]
\centerline{
\includegraphics[scale=0.32,angle=0]{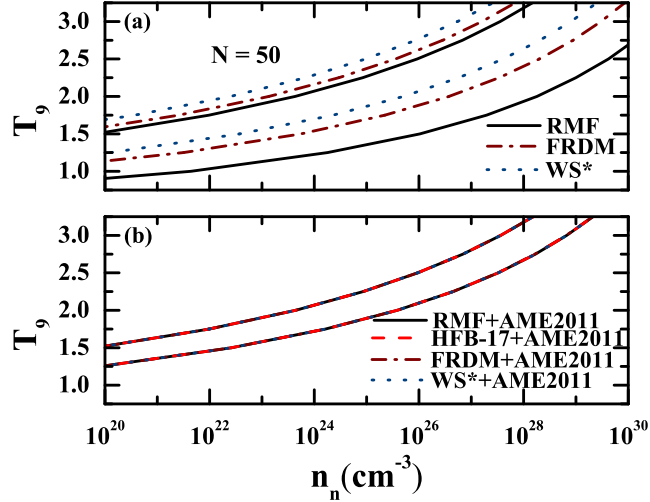}
}
\caption{(Color online) The $T_9$-$n_n$ conditions required by the $N=50$ CWP nuclei.
(a) The band between two curves of the same kind represents the required conditions
based on the corresponding mass model. Note the large differences among the results
for the three indicated models. Note also that no conditions can be found to accommodate
all the $N=50$ CWP nuclei based on the HFB-17 model. (b) Same as (a), but for those
nuclei whose masses are tabulated in the latest atomic mass evaluation
AME2011-preview~\cite{Audi2011}, the model predictions are replaced by the tabulated values.
All four models now give the same conditions required by the $N=50$ CWP nuclei.}
\label{figModel50}
\end{figure}

%----------------------------------------------------------------------------------------
The calculations in Sec.~\ref{sec:nsep} can be generalized to determine
the $T_9$-$n_n$ conditions required by the $N=50$, 82, and 126 CWP
nuclei, respectively, for any specific nuclear mass model. The results for
the $N=50$ CWP nuclei are presented for the FRDM, WS*, and
RMF models in Fig.~\ref{figModel50}(a), which clearly show that the required
conditions change with models. Similar to the case of the $N=82$ CWP
nuclei (Figs.~\ref{figWSpure82} and~\ref{figS2n}) discussed in
Sec.~\ref{sec:nsep}, the upper curve for each model in Fig.~\ref{figModel50}(a)
is effectively determined by the two-neutron separation energy of the lightest
$N=50$ CWP nucleus $^{78}$Ni and the lower curve by that of the $N=52$
isotope $^{82}$Zn of the heaviest $N=50$ CWP nucleus $^{80}$Zn.
Therefore, the large differences among the conditions required by the $N=50$
CWP nuclei for different models can be traced to the differences in the
two-neutron separation energies of $^{78}$Ni and $^{82}$Zn provided by
these models. In particular, the differences for $^{82}$Zn among the models
appear to be substantially larger than those for $^{78}$Ni. We also note that
no conditions can be found to accommodate all the $N=50$ CWP nuclei for
the HFB-17 model, for which the odd-even effects in the neutron separation
energy for Ni, Cu, and Zn isotopes around $N = 50$ are larger by $\sim1$--1.5~MeV
than those for the FRDM, WS*, and RMF models.

%---------------------------------------------------------------------------------------
%----------------------------------------------------------------------------------------

Noting that the two-neutron separation energies of $^{78}$Ni and $^{82}$Zn can be
calculated from the masses tabulated in the latest atomic mass evaluation
AME2011-preview~\cite{Audi2011}, we augment the nuclear mass models by using
the tabulated values in AME2011-preview when they are available to replace the
corresponding model predictions. Remarkably, all four models, including the HFB-17
model, now give the same conditions required by the $N=50$ CWP nuclei as shown
in Fig.~\ref{figModel50}(b). We find that the changes between
Figs.~\ref{figModel50}(a) and~\ref{figModel50}(b) are caused dominantly by the use of
the tabulated masses of $^{76}$Ni to $^{78}$Ni and $^{78}$Zn to $^{82}$Zn, which
confirms the crucial roles of the two-neutron separation energies of $^{78}$Ni and
$^{82}$Zn in determining the conditions required by the $N=50$ CWP nuclei.
In the calculations below, we use the FRDM, WS*, HBF-17, and RMF models
that are augmented by AME2011-preview.

\begin{figure}[h]
\centerline{
\includegraphics[scale=0.48,angle=0]{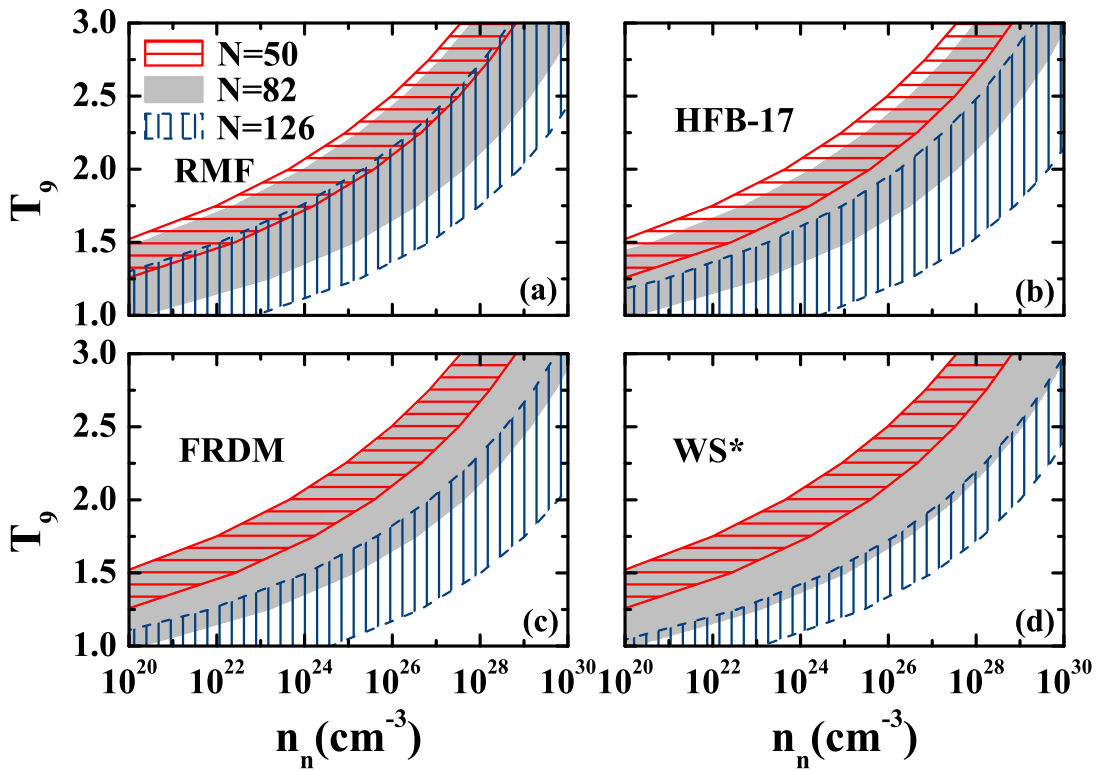}
}
\caption{(Color online) The $T_9$-$n_n$ conditions required by
the $N = 50$ (horizontally hatched band between solid curves), 82 (shaded band), and
126 (vertically hatched band between dashed curves) CWP nuclei, respectively, for four
nuclear mass models: (a) RMF, (b) HFB-17, (c) FRDM, and (d) WS*. See text for details.}
\label{figModelall}
\end{figure}

We calculate the conditions required by the $N=82$ and 126 CWP nuclei, respectively,
as in the case of the $N=50$ CWP nuclei. The results are summarized in Fig.~\ref{figModelall}.
It can be seen that the conditions required by the $N=82$ CWP nuclei (shaded band) are
essentially converged for the four augmented nuclear mass models just like those required by
the $N=50$ CWP nuclei (horizontally hatched band between solid curves).
In contrast, the conditions required
by the $N=126$ CWP nuclei (vertically hatched band between dashed curves)
are still strongly dependent on models. This is because these nuclei
and the majority of those in the nearby region of the nuclear chart are still out of the reach
of experiments while theoretical predictions for their masses involve dramatic extrapolations
with large uncertainties (e.g.,~\cite{Sun2008b}).

%-------------------------------------------------------------------------------------------
%-----------------------------------------------------------------------------------------

Figure~\ref{figModelall} resembles a phase diagram in terms of three bands
for the production of the peaks at
$A\sim 80$, 130, and 195 in $r$-patterns that correspond to the $N=50$, 82, and 126 CWP nuclei.
For the $T_9$-$n_n$ conditions inside the nonoverlap region of a band,
only a single peak can be produced. For those
conditions inside the overlap region of two bands, it is possible to produce two peaks simultaneously.
For the RMF model only, there is a very thin sliver where three bands overlap.
Consequently, we consider it very unlikely that three peaks can be produced simultaneously.
A close examination of Fig.~\ref{figModelall} shows that the $T_9$-$n_n$ conditions required by
the $N = 50$ CWP nuclei (horizontally hatched band between solid curves) are distinct from
those required by the $N=126$ CWP nuclei (vertically hatched band between dashed curves)
for the FRDM, HFB-17, and WS* models. These two sets of conditions overlap only slightly
for the RMF model. This suggests that the peaks at $A\sim 80$ and 195 in $r$-patterns
are not produced simultaneously. Their production may differ in the time of occurrence
within the same astrophysical site or in the astrophysical site itself. In contrast, there is large
overlap between the conditions required by the $N = 50$ and 82 CWP nuclei (shaded band)
for the four models considered. In addition, there is slight to significant overlap between the
conditions required by the $N=82$ and 126 CWP nuclei for all the models.
Therefore, it is possible to produce the peaks at $A\sim 80$ and 130 or those at $A\sim 130$
and 195 simultaneously. The above results will be examined by detailed $r$-process calculations
in Sec.~\ref{sec:patterns}.

%------------------------------------------------------------------------------------------------
\section{$r$-Patterns from the Classical Approach}\label{sec:patterns}
Our main goal here is to explore the effects of nuclear masses on the conditions required
for $r$-process nucleosynthesis under the WP approximation. As discussed in
Sec.~\ref{sec:model}, these conditions are mostly set by the neutron separation energies,
which we calculate from four nuclear mass models augmented by the latest atomic mass
evaluation AME2011-preview. As confirmation of these results, we calculate the
$r$-patterns produced under the conditions shown in Fig.~\ref{figModelall} and compare
them with those inferred for the solar system and observed in metal-poor stars.
As the range of conditions shown in Fig.~\ref{figModelall} is rather broad,
we sample these conditions at a fixed temperature. Specifically, we use $T_9=1.5$ and
$n_n=10^{20.0}$--$10^{22.5}$, $10^{20.5}$--$10^{25.0}$, and $10^{23.5}$--$10^{27.5}$~cm$^{-3}$
as typical conditions required by the $N=50$, 82, and 126 CWP nuclei, respectively.
As noted in Sec.~\ref{sec:cond}, there is overlap between these sets of conditions, which
can lead to coproduction of two peaks in the $r$-pattern.

We carry out an $r$-process calculation using the classical approach (e.g.,~\cite{Cowan1991,Qian2003,Arnould2007})
as follows. We take the seed nucleus to be $^{56}$Fe ($Z=26$). We assume that some material
with an initial abundance $Y_{\rm Fe}$ of $^{56}$Fe is irradiated with neutrons at fixed $T$ and
$n_n$ for a time $\tau$. We then use Eq.~(\ref{equilibrium2}) along with a nuclear mass model to
calculate the fractional abundance $P(Z,A)$ for all the nuclei in each of the isotopic chains with
$Z\geqslant 26$, where
\begin{equation}
P(Z,A)\equiv\frac{Y(Z,A)}{\sum_AY(Z,A)}.
\end{equation}
Note that for fixed $T$ and $n_n$, $P(Z,A)$ is also fixed.
Using the fractional abundances, we calculate the effective $\beta$-decay rate of
an isotopic chain as
\begin{equation}
\lambda_{\beta,Z}\equiv\sum_AP(Z,A)\lambda_\beta(Z,A),
\end{equation}
where $\lambda_\beta(Z,A)$ is the $\beta$-decay rate of the nucleus $(Z,A)$.
For all our calculations, we use
$\lambda_\beta(Z,A)$ from the experimental data in Ref.~\cite{Audi2003} for the nuclei
with measurements and from the theoretical estimates in Ref.~\cite{Moller2003} based
on the FRDM+QRPA method for those without. We then solve the set of equations
\begin{eqnarray}
\dot Y_Z(t)&=&-\lambda_{\beta,Z}Y_Z(t),\ Z=26,\\
\dot Y_Z(t)&=&\lambda_{\beta,Z-1}Y_{Z-1}(t)-\lambda_{\beta,Z}Y_Z(t),\ Z>26,
\end{eqnarray}
where the dot denotes the derivative with respect to time $t$, and $Y_Z(t)$ is the total
abundance of the isotopic chain with proton number $Z$ at time $t$.
The initial conditions are $Y_Z(0)=Y_{\rm Fe}$ for $Z=26$ and 0 for $Z>26$.
We assume that the $r$-process freezes out instantaneously at $t=\tau$. The
freeze-out abundance of the nucleus $(Z,A)$ is
\begin{equation}
Y_{\rm fo}(Z,A)=P(Z,A)Y_Z(\tau).
\end{equation}
The final abundance distribution from an $r$-process episode is obtained by following
the $\beta$ and $\alpha$ decays of all the nuclei in the freeze-out distribution.
We include $\beta$-delayed emission of up to three neutrons~\cite{Moller2003}, which has
the important effect of smoothing the final $r$-pattern. The data on $\alpha$-decays are
taken from the National Nuclear Data Center~\cite{NNDC}. Fission is ignored in all the calculations.
Note that the WP approximation is implicit in the classical approach as $Y_Z(t)$ is
dominated by the corresponding WP nucleus with $P(Z,A_{\rm WP})\geqslant 0.5$.

For comparison with the $r$-patterns inferred for the solar system and observed in
metal-poor stars, we need to superpose the results from many $r$-process episodes
described above. As we take $T_9=1.5$ for all the calculations, we denote each
episode by its $n_n$. The neutron irradiation time $\tau(n_n)$ and the weight
$\omega(n_n)$ for each episode are taken to be
\begin{eqnarray}\label{tau_weight}
\tau(n_n) &=& a\times n_{n}^{b},\\
\omega(n_n) &=& c\times n_{n}^{d},
\end{eqnarray}
where $a$, $b$, $c$, and $d$ are parameters to be determined by a least-squares fit to the
$r$-pattern used for comparison. While such a superposition procedure is a crude
approximation to estimate $r$-patterns produced by astrophysical sources, it can still provide
some useful information on the conditions that the actual $r$-process sites must fulfill~\cite{Kratz2007}.
For this reason, this procedure has been used extensively in $r$-process studies (e.g.,~\cite{Kratz1993,Freiburghaus1999b,Schatz2002,Kratz2007,Sun2008a,Niu2009,Li2012}).

\subsection{Comparison with solar-like $r$-patterns}

\begin{figure}[h]
\centerline{
\includegraphics[scale=0.48,angle=0]{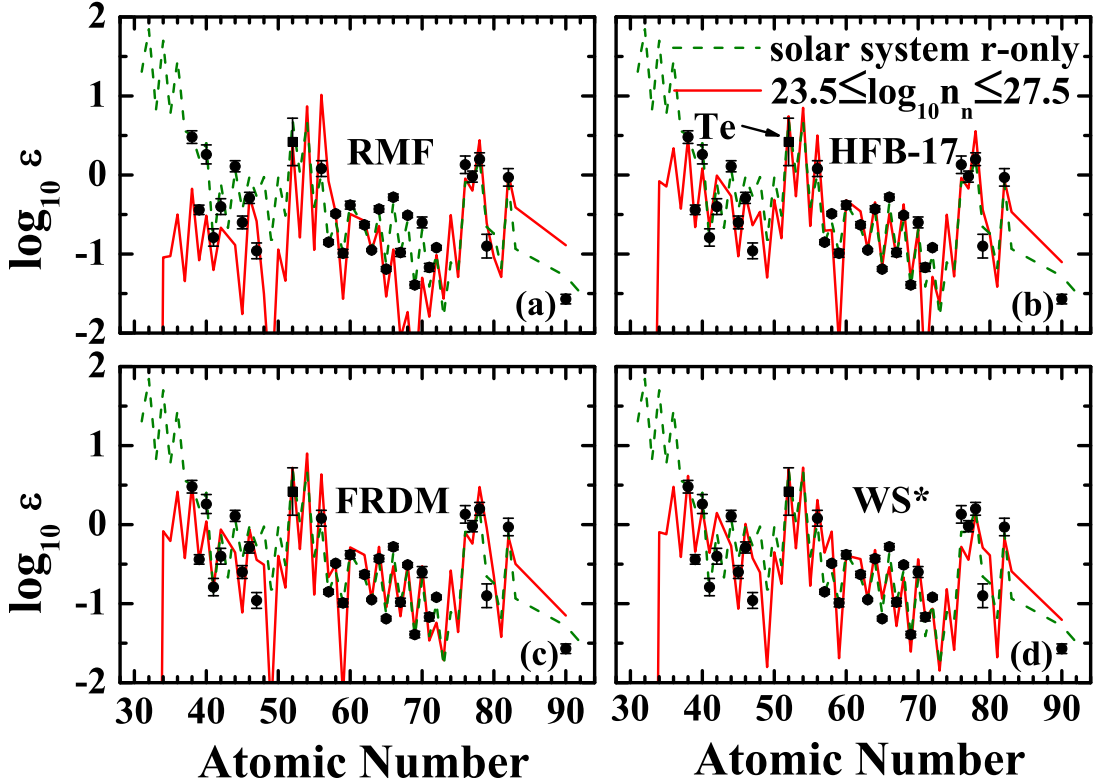}
} \caption{(Color online) Elemental abundances (solid curves) calculated from the classical
$r$-process approach using four nuclear mass models: (a) RMF, (b) HFB-17, (c) FRDM,
and (d) WS*. The filled circles represent the average data on $r$-II stars of the Milky Way
halo and the dashed curve is the solar $r$-pattern translated to pass through the Eu data.
The filled square gives the Te abundance recently measured in the metal-poor star
BD~+17$^{\circ}$3248 (shifted according to the observed Te/Eu ratio). The calculated solid curves
(Fit I) are the best-fit results to reproduce the solar isotopic $r$-pattern for $125\leqslant A\leqslant209$
with $T_9=1.5$ and $10^{23.5}\leqslant n_n\leqslant 10^{27.5}$~cm$^{-3}$, which are the typical
conditions required by the $N=126$ CWP nuclei. See text for details.}
\label{figFitSolar1}
\end{figure}

We first use the classical approach to reproduce the solar $r$-pattern for $125\leqslant A\leqslant209$~\cite{Cowan2006}
(see also, e.g.,~\cite{Sneden2003,Niu2009}), which is shown as the dashed
curve in Fig.~\ref{figFitSolar1}. We consider a superposition of nine neutron
densities (equidistant on a $\log_{10}$ scale) within the range
$10^{23.5}\leqslant n_n\leqslant 10^{27.5}$~cm$^{-3}$, which corresponds to the typical conditions
required by the $N=126$ CWP nuclei for $T_9=1.5$ (see Fig.~\ref{figModelall}).
The best-fit results (hereafter ``Fit I'') for the four adopted nuclear mass models are shown as
the solid curves in Fig.~\ref{figFitSolar1}. Note that although the fits are performed for the
solar isotopic $r$-pattern, the patterns shown in Fig.~\ref{figFitSolar1} are for the
corresponding elemental abundances. It can be seen that the solar $r$-pattern
from the peak at $A\sim 130$ ($Z\sim 52$, Te) to that at $A\sim 195$ ($Z\sim 78$, Pt)
is reproduced rather well for the FRDM, HBF-17, and WS* models. For the RMF model,
the rare-earth elements with $Z=66$--70 (Dy, Ho, Er, Tm, and Yb) are severely underproduced.
This deficiency may reflect the necessity to adopt improved RMF parameter sets
(e.g., PC-PK1~\cite{Zhao2010}, which provides a much better description for the
properties of nuclear ground and excited states~\cite{Meng2006})
or that the classical approach is inadequate to give a full description of $r$-process nucleosynthesis
(e.g., instantaneous freeze-out is not a good approximation~\cite{Freiburghaus1999b}).
In any case, the peaks at $A\sim 130$ and 195 are reproduced adequately for all four mass models,
which suggests that the typical conditions required by the $N=126$ CWP nuclei can indeed produce both
these peaks. As discussed in Sec.~\ref{sec:cond}, this is because there is significant overlap
between the conditions required by the $N=82$ and 126 CWP nuclei (see Fig.~\ref{figModelall}).

%------------------------------------------------------------------------------------------------
%------------------------------------------------------------------------------------------------

Observations show that Ba ($Z=56$) and heavier elements in many metal-poor stars
of the Milky Way halo follow the solar $r$-pattern rather closely~\cite{Sneden2008}.
The values of $\log_{10}\varepsilon({\rm E})\equiv\log_{10}({\rm N_E/N_H})+12$, where $N_E$ and $N_H$
represent the abundance of element E and hydrogen respectively, obtained for the elements with $38\leqslant Z\leqslant 79$
by averaging the data~\cite{Sneden2008,Sneden2003,Hill2002,Sneden2009} on two
such ``$r$-II'' stars, CS~22892--052 and CS~31082--001, are shown as filled circles in Fig.~\ref{figFitSolar1}.
The data on Pb ($Z=82$) and Th ($Z=90$) for CS~22892--052 are also shown. The solar $r$-pattern has been
translated to pass through the data on Eu ($Z=63$) and can be seen to represent
the pattern for $Z\geqslant 56$ in $r$-II stars very well.
Recently, Te ($Z=52$) has been measured in a group of
metal-poor stars (BD~+17$^{\circ}$3248, HD~108317, and HD~128279~\cite{Roederer2012a} and
HD~160617~\cite{Roederer2012b}). This extends the comparison of
$r$-patterns in metal-poor stars with the solar $r$-pattern to include an
element in the peak at $A\sim130$. The data on Te for BD~+17$^{\circ}$3248 (shifted
according to the observed Te/Eu ratio) is shown as the filled square in Fig.~\ref{figFitSolar1}.
It can be seen that the Te data is consistent with the solar $r$-pattern and with
the coproduction of the peaks at $A\sim 130$ and 195 under the typical conditions required
by the $N=126$ CWP nuclei. The above results are also in agreement with previous
studies (e.g.,~\cite{Kratz2007,Niu2009}), in which it was concluded that the $r$-process responsible
for the elements with $56\leqslant Z\leqslant82$ is characterized by neutron densities of
$10^{23}$--$10^{28}$~cm$^{-3}$.

\begin{figure}[h]
\centerline{
\includegraphics[scale=0.48,angle=0]{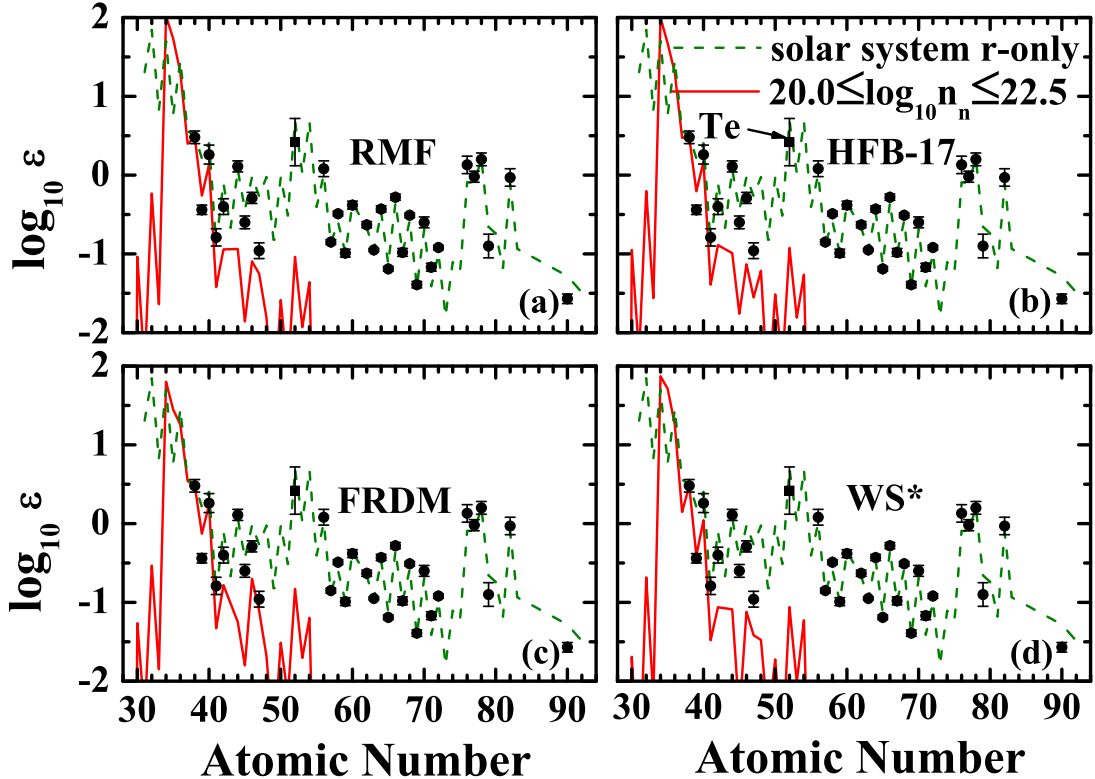}
} \caption{(Color online) Same as Fig.~\ref{figFitSolar1}, but the solid curves (Fit II)
are the best-fit results to reproduce the solar isotopic $r$-pattern for $69\leqslant A\leqslant124$
with $T_9=1.5$ and $10^{20.0}\leqslant n_n\leqslant 10^{22.5}$~cm$^{-3}$, which are the typical
conditions required by the $N=50$ CWP nuclei. See text for details.}
\label{figFitSolar2}
\end{figure}

The Fit I results shown in Fig.~\ref{figFitSolar1} cannot adequately reproduce the
abundances of the elements with $38\leqslant Z\leqslant 47$ in $r$-II stars
(especially when the RMF model is used).
Further, the elements in the peak at $A\sim 80$ ($Z\sim 34$)
of the solar $r$-pattern are severely underproduced by these calculations.
Additional $r$-process contributions to or alternative sources for the elements below
the peak at $A\sim 130$ are thus required and this issue has been under active
investigation~\cite{Qian2001,Travaglio2004,Montes2007,Ishimaru2005,Qian2007,Arcones2011b}.
Here we explore the possibility that there are additional contributions from $r$-process
nucleosynthesis under the conditions required by the $N=50$ CWP nuclei.
We consider a superposition of six neutron densities (equidistant on a $\log_{10}$ scale)
within the range $10^{20.0}\leqslant n_n\leqslant 10^{22.5}$~cm$^{-3}$ to best reproduce the solar
isotopic $r$-pattern for $69\leqslant A\leqslant124$ (hereafter ``Fit II''). The results are shown as the
solid curves in Fig.~\ref{figFitSolar2}. It can be seen that the conditions required by the
$N=50$ CWP nuclei indeed can produce the peak at $A\sim 80$. However, it is also clear
that the $r$-patterns from the peak at $A\sim 130$ to that at $A\sim 195$ inferred for
the solar system and observed in $r$-II stars require very different conditions from those
for producing the peak at $A\sim 80$ (see Sec. IV).

%------------------------------------------------------------------------------------------------
%------------------------------------------------------------------------------------------------

\begin{figure}[h]
\centerline{
\includegraphics[scale=0.48,angle=0]{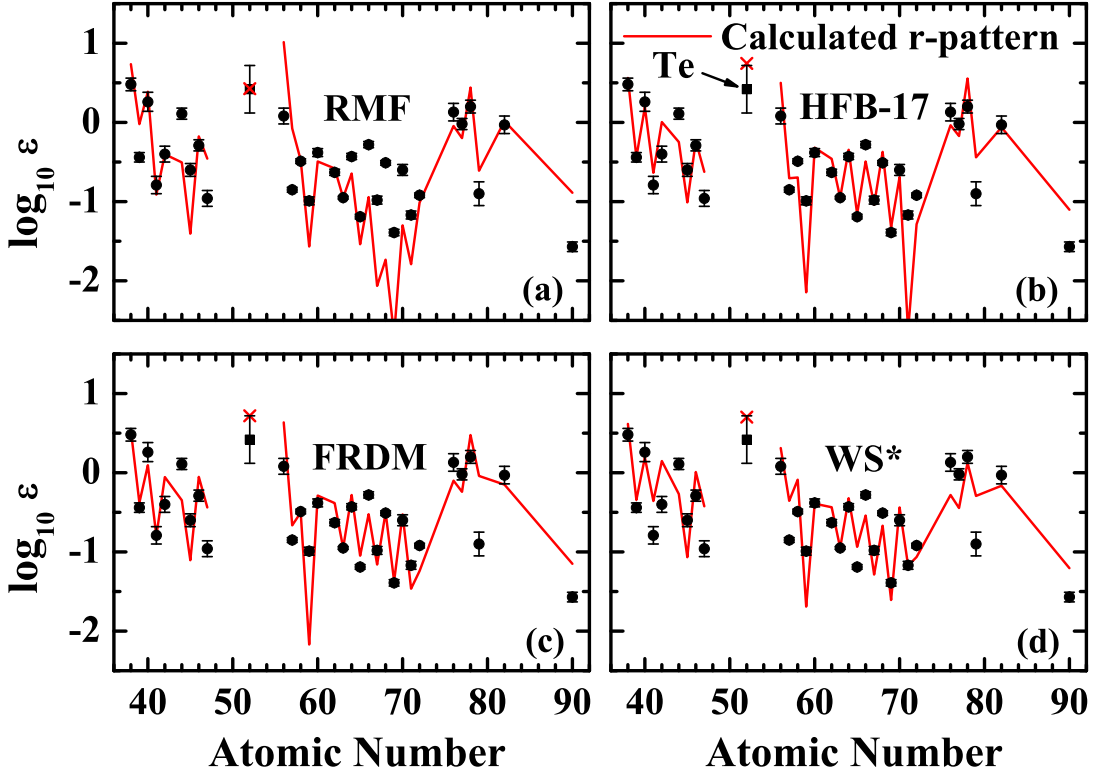}
} \caption{(Color online) Best fits (solid curves) to the $r$-pattern in $r$-II stars
(filled circles) using a superposition of neutron densities within the ranges
 $10^{20.0}\leqslant n_n\leqslant 10^{22.5}$~cm$^{-3}$ and $10^{23.5}\leqslant n_n\leqslant 10^{27.5}$~cm$^{-3}$
for four nuclear mass models: (a) RMF, (b) HFB-17, (c) FRDM, and (d) WS*.
The crosses give the calculated Te abundances, which are consistent with the measured
value (filled square) for the metal-poor star BD~+17$^{\circ}$3248
(shifted according to the observed Te/Eu ratio). See text for details.}
\label{figSumR-II}
\end{figure}

To find the best match to the $r$-pattern in $r$-II stars, we consider a superposition of
neutron densities in the two ranges adopted for Fits I and II. The results are shown as
the solid curves in Fig.~\ref{figSumR-II}. It can be seen that fair agreement between
the calculated and observed patterns is obtained for the FRDM, HFB-17, and
WS* models. Note also that the calculated Te abundances (crosses) for all four models
are consistent with the newly measured value for the metal-poor star BD~+17$^{\circ}$3248.
However, the trough at $Z=66$--70 is clearly problematic for the RMF model.
This may be caused by nuclear shape transition before the $N=126$ closed neutron shell
and the location of the transition region could have been assigned incorrectly in the
RMF model~\cite{Sun2008a}. Discrepancies can also be seen for Ru, Rh, and Ag
($Z=44$, 45, and 47, respectively) for all four models. This issue needs to be addressed
by detailed considerations of the astrophysical environments for the $r$-process and
alternative sources for the elements below the peak
at $A\sim 130$~\cite{Qian2001,Travaglio2004,Montes2007,Ishimaru2005,Qian2007,Arcones2011b}.

%------------------------------------------------------------------------------------------------
%------------------------------------------------------------------------------------------------

%--------------------------------------------------------------------------------------
\subsection{Comparison with a non-solar-like $r$-pattern}
\begin{figure}[h]
\centerline{
\includegraphics[scale=0.48,angle=0]{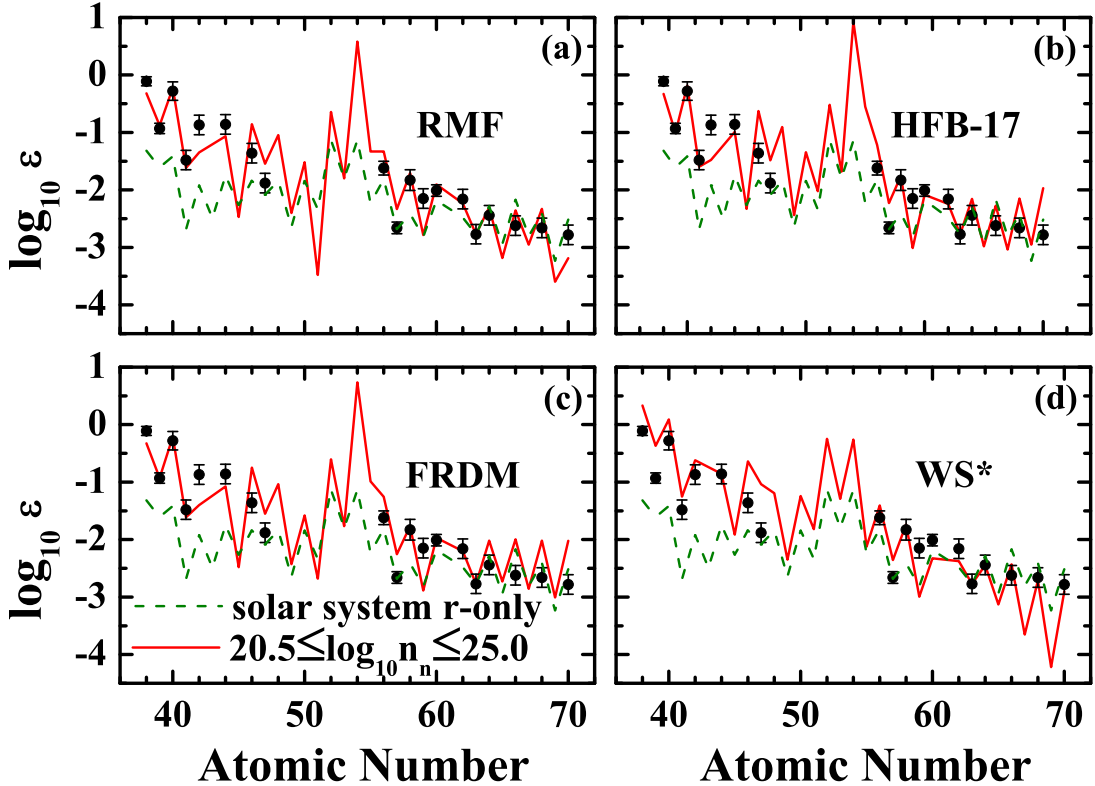}
} \caption{(Color online) Best fits (solid curves) to the non-solar-like $r$-pattern
(filled circles) in the metal-poor star HD~122563
for four nuclear mass models: (a) RMF, (b) HFB-17, (c) FRDM, and (d) WS*.
These results use a superposition of neutron densities within the range
$10^{20.5}\leqslant n_n\leqslant 10^{25.0}$~cm$^{-3}$, which corresponds to the typical
conditions required by the $N=82$ CWP nuclei for $T_9=1.5$.
The dashed curve gives the solar $r$-pattern translated to pass through
the filled circle for Eu ($Z=63$). See text for details.}
\label{figFitHD}
\end{figure}

%-------------------------------------------------------------------------------------
In contrast to the $r$-II stars, some metal-poor stars exhibit
an $r$-pattern that is clearly different from the solar one. Prominent examples are
the metal-poor stars HD~122563~\cite{Honda2006} and HD~88609~\cite{Honda2007},
which have almost the same abundances for Cu ($Z=29$) and heavier elements.
The data on Sr ($Z=38$) and heavier elements for HD~122563 are shown as the filled
circles in Fig.~\ref{figFitHD}. Relative to the solar $r$-pattern translated to pass
through the Eu data (dashed curve), the elements below the peak at $A\sim 130$
in this star are grossly overabundant. It was argued that in addition to an $r$-process
source for producing a solar-like $r$-pattern from the peak at $A\sim 130$ to that at
$A\sim 195$, a very different source is required to explain the data for stars like HD~122563
\cite{Qian2007,Honda2006,Honda2007}.

%-------------------------------------------------------------------------------------------------
%--------------------------------------------------------------------------------------------------

Here we attempt to interpret the abundance pattern observed in HD~122563 using the
classical $r$-process approach. We find that the conditions required by the $N = 82$ CWP
nuclei can best reproduce this pattern while those required by the $N=50$ and 126 CWP
nuclei can not. The best-fit results use a superposition of ten neutron densities
(equidistant on a $\log_{10}$ scale) within the range $10^{20.5}\leqslant n_n\leqslant 10^{25.0}$~cm$^{-3}$,
which corresponds to the typical conditions required by the $N=82$ CWP nuclei for $T_9=1.5$
(see Fig.~\ref{figModelall}). These results are shown as the solid curves in Fig.~\ref{figFitHD}.
It can be seen that an approximate overall match of the calculated with the observed abundances
is obtained for all four nuclear mass models. Therefore, we suggest that it is plausible to account
for the abundance pattern in stars like HD~122563 by an $r$-process operating under the
conditions required by the $N = 82$ CWP nuclei. We note that Fig.~\ref{figFitHD} shows
a clear difference in the calculated relative production of Te and Xe ($Z=52$ and 54, respectively)
between the WS* and the other three mass models: these two elements are produced in
approximately equal amount for the WS* model but Xe is produced much more than Te for the
other three models. Measurements of these two elements in HD~122563 would be extremely
valuable in constraining nuclear mass models although they also represent a difficult challenge
to spectroscopic observations.

%-----------------------------------------------------------------------------------------
\section{Discussion and Conclusions}\label{sec:discuss}
\begin{figure}[h]
\centerline{
\includegraphics[scale=0.48,angle=0]{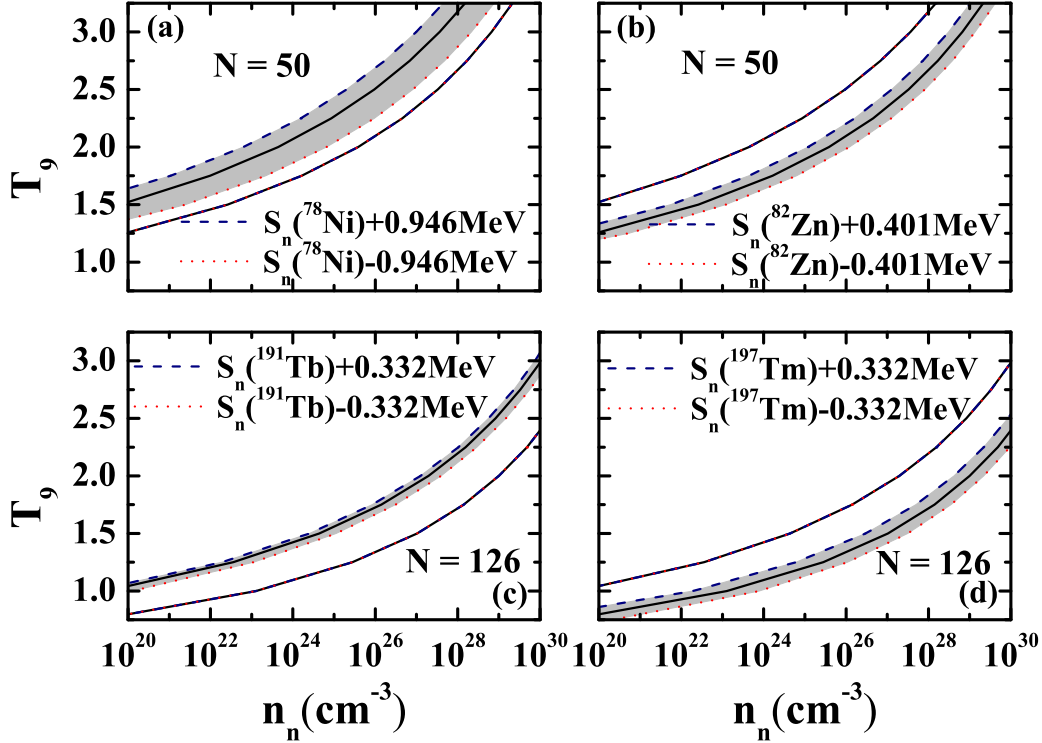}
} \caption{(Color online) Effects of the uncertainty in the neutron separation energy $S_n$
for (a) $^{78}$Ni, (b) $^{82}$Zn, (c) $^{191}$Tb, and (d) $^{197}$Tm on the required $T_9$-$n_n$
conditions. The solid curves in panels (a) and (b) are the same as those in Fig.~\ref{figModel50}(b).
The solid curves in panels (c) and (d) are the same as the dashed curves (for the WS* model) in Fig.~\ref{figModelall}(d).
The shaded regions in each panel show the effects on
the required $T_9$-$n_n$ conditions when the corresponding $S_n$ values are varied
within the estimated uncertainties. See text for details.}
\label{figErr}
\end{figure}

%-----------------------------------------------------------------------------------------
We have explored the effects of four nuclear mass models (FRDM, WS*, HBF-17, and RMF)
on the conditions required by $r$-process nucleosynthesis under the WP approximation.
As discussed in Sec.~\ref{sec:model}, the required $T_9$-$n_n$ conditions are mostly
determined by the two-neutron separation energies of the CWP nuclei with $N=50$, 82,
and 126 and of those nuclei around them. Figure~\ref{figModel50} shows the dramatic effect
of using the tabulated values in the latest atomic mass evaluation AME2011-preview when
they are available to replace the masses predicted by models. As noted in Sec.~\ref{sec:cond},
the tabulated masses of $^{76}$Ni to $^{78}$Ni and $^{78}$Zn to $^{82}$Zn
play crucial roles in determining the conditions required by the $N=50$ CWP nuclei.
However, the tabulated masses of $^{76}$Ni to $^{78}$Ni and $^{82}$Zn are extrapolated
rather than measured. To emphasize the effects of these masses on the conditions required
by the $N=50$ CWP nuclei, we first repeat the calculations of Sec.~\ref{sec:model} by varying
the neutron separation energy of $^{78}$Ni within the estimated uncertainty of 0.946~MeV
\cite{Audi2011} while keeping the other input the same as for Fig.~\ref{figModel50}(b). The
results are shown in Fig.~\ref{figErr}(a). In comparison with Fig.~\ref{figModel50}(b),
the lower bound on the region of the required $T_9$-$n_n$ conditions stays the same
because this is determined by the two-neutron separation energy of $^{82}$Zn
(see Sec.~\ref{sec:cond}), which is not changed. Increasing the neutron separation energy of
$^{78}$Ni by 0.946~MeV raises the upper bound from
the solid curve [upper bound in Fig.~\ref{figModel50}(b)] to the dashed curve and
decreasing this quantity by the same amount lowers it to the dotted curve.
We then repeat the same calculations but vary the neutron separation energy of
$^{82}$Zn within the estimated uncertainty of 0.401~MeV~\cite{Audi2011}.
The effects on the lower bound on the region of the required $T_9$-$n_n$ conditions are
shown in Fig.~\ref{figErr}(b). Note that if the neutron separation energy of $^{78}$Ni were
lower than its tabulated value by 0.946~MeV while that of $^{82}$Zn were higher by 0.401~MeV,
then it would be almost impossible to find any $T_9$-$n_n$ conditions to accommodate all the
$N=50$ CWP nuclei. In any case, the significant effects of uncertainties in neutron separation
energies on the required $T_9$-$n_n$ conditions shown in panels (a) and (b) of Fig.~\ref{figErr}
clearly demonstrate the importance of precise mass measurements for $^{76}$Ni to $^{78}$Ni and $^{82}$Zn.

%-----------------------------------------------------------------------------------------
%-------------------------------------------------------------------------------------------

As in the case of $N=50$ CWP nuclei, we have also made a careful scan of the nuclear chart
around the $N = 82$ CWP nuclei and explored the effects of those nuclei with experimentally
unknown or poorly measured masses on the $T_9$-$n_n$ conditions required by the $N = 82$ CWP nuclei.
When nuclear masses are not known experimentally, we have used the extrapolated masses
and uncertainties as these have been proven to have a better predictive power than all
available models~\cite{Fu2011,Sun2011}. We have identified $^{131}$Cd and $^{132}$Cd as
additional key nuclei for precise mass measurements.

As can be seen from Fig.~\ref{figModelall}, the $T_9$-$n_n$ conditions required by the
$N=126$ CWP nuclei depend strongly on the nuclear mass model. To assess the impact of
uncertainties in nuclear mass models, we varied the neutron separation energies $S_n$
for the relevant nuclei within the known errors or the rms deviations of model predictions
for the known masses. Using the WS* model as an example, we show the effects of uncertainties
in $S_n$ for $^{191}$Tb and $^{197}$Tm on the required $T_9$-$n_n$ conditions in panels
(c) and (d) of Fig.~\ref{figErr}, respectively. It can be seen that the upper bound on these
conditions changes very little when $S_n(^{191}{\rm Tb})$ is varied within 0.332 MeV
(1 rms deviation for the WS* model) but the lower bound is more sensitive to the same
variation of $S_n(^{197}{\rm Tm})$. However, the conditions required by the $N=50$
[Figs.~\ref{figErr}(a) and 9(b)] and 126 [Figs.~\ref{figErr}(c) and 9(d)] CWP nuclei do not appear
to overlap for the WS* model even when uncertainties in nuclear masses are considered.
This suggests that, at least for this model, the $A\sim 80$ and 195 peaks in the
$r$-pattern are most likely produced under very different conditions.
Precise mass measurements and better calibrated mass models are needed to make this result more robust.

In conclusion, we have estimated the temperature and neutron density conditions required
for $r$-process nucleosynthesis under the WP approximation using four nuclear mass models
augmented by the latest atomic mass evaluation AME2011-preview.
We have shown that these conditions are mostly determined by the two-neutron separation
energies of the $N=50$, 82, and 126 CWP nuclei and those around them. We have also
identified some key nuclei including $^{76}$Ni to $^{78}$Ni, $^{82}$Zn, $^{131}$Cd, and
$^{132}$Cd for precise mass measurements at rare-isotope beam facilities.

Based on the typical conditions required by the $N=50$, 82, and 126 CWP nuclei shown
in Fig.~\ref{figModelall}, we have performed $r$-process calculations in the classical
approach to reproduce the $r$-pattern inferred for the solar system and those observed
in metal-poor stars of the Milky Way halo. We have found that (1) at least for
the nuclear mass models considered here, the conditions required
to produce the peak at $A\sim 80$ differ greatly from those required to produce the
solar $r$-pattern from the peak at $A\sim 130$ to that at $A\sim 195$, which reflects that
the $T_9$-$n_n$ conditions required by the $N=50$ and 126 CWP nuclei are very
different (especially for the WS* model); (2) the solar $r$-pattern from the peak at
$A\sim 130$ to that at $A\sim 195$, which also closely describes the $r$-patterns in many
metal-poor stars, can be reproduced under the conditions required by the $N=126$ CWP nuclei, which
has significant overlap with those required by the $N=82$ CWP nuclei, thereby enabling
coproduction of the peaks at $A\sim 130$ and 195; (3) it is plausible to explain the
overall $r$-patterns in metal-poor $r$-II stars with a superposition of two sets of $r$-process
conditions required by the $N = 50$ and 126 CWP nuclei, respectively; and
(4) the non-solar-like $r$-pattern observed in metal-poor stars like HD~122563 can be
accounted for by the $r$-process conditions required by the $N = 82$ CWP nuclei.
We note that similar results were also obtained by other earlier studies (e.g.,~\cite{Kratz2007,Schatz2002}).

We recognize that the classical $r$-process approach leaves out many important
details, such as the time evolution of temperature and neutron density, the finite duration
of the freeze-out, and the breakdown of $(n,\gamma)\rightleftharpoons(\gamma,n)$
equilibrium during the freeze-out. We note that the impact of the details of the
freeze-out on the final $r$-pattern~\cite{Surman2009,Arcones2011a}, especially the formation of the rare-earth peak~\cite{Mumpower2012a,Mumpower2012b}, has been investigated extensively in other recent studies. However, so long as
$(n,\gamma)\rightleftharpoons(\gamma,n)$ equilibrium can be achieved in an $r$-process environment,
the conditions immediately before the freeze-out in that environment should be close to those derived here.
 We intend to carry out parametric studies of the $r$-process based on
more detailed and more realistic astrophysical models in the future, and will explore
the effects of various nuclear input on such models.

%=================================================================
\acknowledgments This work was supported in part by the 973 Program
(Grant No. 2013CB834400), the National Natural Science Foundation of
China (Grants No. 10975007, No. 10975008, No. 11005069, No. 11035007, No.
11105010, No. 11128510, No. 11175002, and No. 11205004.), the
Research Fund for the Doctoral Program of Higher Education (Grant No.
20110001110087), the Program for New Century Excellent Talents in
University (Grant No. NCET-09-0031), and the 211 Project of Anhui
University (Grant No. 02303319-33190135) in the People's Republic of
China and by the U.S. Department of Energy under Grant No.
DE-FG02-87ER40328 at the University of Minnesota.
%=================================================================

%------------------------------------------------------------------
%\clearpage
%\bibliographystyle{apsrev}

%\bibliography{reference} % Produces the bibliography via BibTeX.

\begin{thebibliography}{75}

\bibitem{Burbidge1957}
E.~M. Burbidge, G.~R. Burbidge, W.~A. Fowler, and F. Hoyle, Rev. Mod. Phys. {\bf 29}, 547 (1957).

\bibitem{Cameron1957}
A.~G. W. Cameron, Chalk River Report {\bf CRL-41}, 1957.

\bibitem{Cowan1991}
J.~J. Cowan, F.-K. Thielemann, and J.~W. Truran, Phys. Rep. {\bf 208}, 267 (1991).

\bibitem{Qian2003}
Y.-Z. Qian, Prog. Part. Nucl. Phys. {\bf 50}, 153 (2003).

\bibitem{Arnould2007}
M. Arnould, S. Goriely, and K. Takahashi, Phys. Rep. {\bf 450}, 97 (2007).

\bibitem{Kappeler2011}
F. K\"appeler, R. Gallino, S. Bisterzo, and W. Aoki, Rev. Mod. Phys. {\bf 83}, 157 (2011).

\bibitem{Woosley1992}
S.~E. Woosley and R.~D. Hoffman, Astrophys. J. {\bf 395}, 202 (1992).

\bibitem{Meyer1992}
B.~S. Meyer, G.~J. Mathews, W.~M. Howard, S.~E. Woosley, and R.~D. Hoffman, Astrophys. J. {\bf 399}, 656 (1992).

\bibitem{Woosley1994}
S.~E. Woosley, J.~R. Wilson, G.~J. Mathews, R.~D. Hoffman, and B.~S. Meyer, Astrophys. J. {\bf 433}, 229 (1994).

\bibitem{Ning2007}
H. Ning, Y.-Z. Qian, and B.~S. Meyer, Astrophys. J. {\bf 667}, L159 (2007).

\bibitem{Pruet2003}
J. Pruet, S.~E. Woosley, and R.~D. Hoffman, Astrophys. J. {\bf 586}, 1254 (2003).

\bibitem{Surman2006}
R. Surman, G.~C. McLaughlin, and W.~R. Hix, Astrophys. J. {\bf 643}, 1057 (2006).

\bibitem{Surman2008}
R. Surman, G.~C. McLaughlin, M. Ruffert, H.-T. Janka, and W.~R. Hix, Astrophys. J. {\bf 679}, L117 (2008).

\bibitem{Wanajo2012}
S. Wanajo and H.-T. Janka, Astrophys. J. {\bf 746}, 180 (2012).

\bibitem{Epstein1988}
R.~I. Epstein, S.~A. Colgate, and W.~C. Haxton, Phys. Rev. Lett. {\bf 61}, 2038 (1988).

\bibitem{Banerjee2011}
P. Banerjee, W.~C. Haxton, and Y.-Z. Qian, Phys. Rev. Lett. {\bf 106}, 201104 (2011).

\bibitem{Lattimer1977}
J.~M. Lattimer, F. Mackie, D.~G. Ravenhall, and D.~N. Schramm, Astrophys. J. {\bf 213}, 225 (1977).

\bibitem{Freiburghaus1999a}
C. Freiburghaus, S. Rosswog, and F.-K. Thielemann, Astrophys. J. {\bf 525}, L121 (1999).

\bibitem{Goriely2011}
S. Goriely, A. Bauswein, and H.-T. Janka, Astrophys. J. {\bf 738}, L32 (2011).

\bibitem{Korobkin2012}
O. Korobkin, S. Rosswog, A. Arcones, and C. Winteler,  Mon. Not. R. Astron. Soc. {\bf 426}, 1940 (2012).

\bibitem{Janka2012}
H.-T. Janka, Annu. Rev. Nucl. Part. Sci. 62, 407 (2012).

\bibitem{Martinez2012}
G. Mart\'inez-Pinedo, T. Fischer, A. Lohs, and L. Huther, arXiv:1205.2793.

\bibitem{Roberts2012}
L.~F. Roberts, S. Reddy, and G. Shen, Phys. Rev. C {\bf 86}, 065803 (2012).

\bibitem{Qian2000}
Y.-Z. Qian, Astrophys. J. {\bf 534}, L67 (2000).

\bibitem{Argast2004}
D. Argast, M. Samland, F.-K. Thielemann, and Y.-Z. Qian, Astron. Astrophys. {\bf 416}, 997 (2004).

\bibitem{Donder2004}
E.~De Donder and D. Vanbeveren, New Astron. Rev. {\bf 48}, 861 (2004).

\bibitem{Page2006}
D. Page and S. Reddy, Annu. Rev. Nucl. Part. Sci. {\bf 56}, 327 (2006).

\bibitem{Freiburghaus1999b}
C. Freiburghaus, J.-F. Rembges, T. Rauscher, E. Kolbe, F.-K. Thielemann, K.-L. Kratz, B. Pfeiffer,
and J.~J. Cowan, Astrophys. J. {\bf 516}, 381 (1999).

\bibitem{Goriely1996}
S. Goriely and M. Arnould, Astron. Astrophys. {\bf 312}, 327 (1996).

\bibitem{Nishimura2011}
S. Nishimura {\it et al.}, Phys. Rev. Lett. {\bf 106}, 052502 (2011).

\bibitem{Baruah2008}
S. Baruah {\it et al.}, Phys. Rev. Lett. {\bf 101}, 262501 (2008).

\bibitem{Sun2008b}
B. Sun {\it et al}, Nucl. Phys. A {\bf 812}, 1 (2008).

\bibitem{Dillmann2003}
I. Dillmann {\it et al.}, Phys. Rev. Lett. {\bf 91}, 162503 (2003).

\bibitem{Audi2011}
G. Audi and W. Meng (private communication); http://amdc.in2p3.fr/masstables/Ame2011int/filel.html.

\bibitem{Moller1995}
P. M\"oller, J. Nix, W. Myers, and W. Swiatecki, At. Data Nucl. Data Tables {\bf 59}, 185 (1995).

\bibitem{Wang2010}
N. Wang, Z. Liang, M. Liu, and X. Wu, Phys. Rev. C {\bf 82}, 044304 (2010).

\bibitem{Goriely2009}
S. Goriely, N. Chamel, and J.~M. Pearson, Phys. Rev. Lett. {\bf 102}, 152503 (2009).

\bibitem{Geng2005}
L. Geng, H. Toki, and J. Meng, Prog. Theor. Phys. {\bf 113}, 785 (2005).

\bibitem{Sneden2008}
C. Sneden, J.~J. Cowan, and R. Gallino, Annu. Rev. Astron. Astrophys. {\bf 46}, 241 (2008).

\bibitem{Qian2007}
Y.-Z. Qian and G.Wasserburg, Phys. Rep. {\bf 442}, 237 (2007).

\bibitem{Farouqi2009}
K. Farouqi, K.-L. Kratz, L.~I. Mashonkina, B. Pfeiffer, J.~J. Cowan, F.-K. Thielemann, and
J.~W. Truran, Astrophys. J. {\bf 694}, L49 (2009).

\bibitem{Farouqi2010}
K. Farouqi, K.-L. Kratz, B. Pfeiffer, T. Rauscher, F.-K. Thielemann, and J.~W. Truran,
Astrophys. J. {\bf 712}, 1359 (2010).

\bibitem{Kratz1993}
K.-L. Kratz, J.~P. Bitouzet, F.~K. Thielemann, P. M\"oller, and B. Pfeiffer,
Astrophys. J. {\bf 403}, 216 (1993).

\bibitem{Surman2009}
R. Surman, J. Beun, G.~C. McLaughlin, and  W.~R. Hix, Phys. Rev. C {\bf 79}, 045809 (2009).

\bibitem{Arcones2011a}
A. Arcones, and G. Mart\'inez-Pinedo, Phys. Rev. C {\bf 83}, 045809 (2011).

\bibitem{Arcones2012}
A. Arcones, and G.~F. Bertsch, Phys. Rev. Lett. {\bf 108}, 151101 (2012).

\bibitem{Kratz1988}
K.-L. Kratz, F.~K. Thielemann, W. Willebrandt, P. M\"oller, V. Harms, A.Wohr, and J.~W. Truran,
J. Phys. G {\bf 14}, S331 (1988).

\bibitem{Goriely1992}
S. Goriely and M. Arnould, Astron. Astrophys. {\bf 262}, 73 (1992).

\bibitem{Audi2003}
G. Audi, O. Bersillon, J. Blachot, and A. H. Wapstra, Nucl. Phys. A {\bf 729}, 3 (2003).

\bibitem{Moller2003}
P. M\"oller, B. Pfeiffer, and K.-L. Kratz, Phys. Rev. C {\bf 67}, 055802 (2003).

\bibitem{NNDC}
National Nuclear Data Center, [http://www.nndc.bnl.gov].

\bibitem{Kratz2007}
K.-L. Kratz, K. Farouqi, B. Pfeiffer, J.~W. Truran, C. Sneden, and J.~J. Cowan,
Astrophys. J. {\bf 662}, 39 (2007).

\bibitem{Schatz2002}
H. Schatz, R. Toenjes, B. Pfeiffer, T.~C. Beers,2 J.~J. Cowan, V. Hill, and K.-L. Kratz,
Astrophys. J. {\bf 579}, 626 (2002).

\bibitem{Sun2008a}
B. Sun, F. Montes, L.~S. Geng, H. Geissel, Y.~A. Litvinov, and J. Meng,
Phys. Rev. C {\bf 78}, 025806 (2008).

\bibitem{Niu2009}
Z. Niu, B. Sun, and J. Meng, Phys. Rev. C {\bf 80}, 065806 (2009).

\bibitem{Li2012}
Z. Li, Z.~M. Niu, B. Sun, N. Wang, and J. Meng, Acta Phys. Sin. {\bf 61},
072601 (2012) (in Chinese).

\bibitem{Cowan2006}
J.~J. Cowan, J.~E. Lawler, C. Sneden, E.~A. den Hartog, and J. Collier,
in \emph{Proceedings of the 2006 NASA Laboratory Astrophysics Workshop} (NASA/CP-2006-214549),
edited by V. H. S. Kwong and F. S. Wreck (NASA Center for Aerospace
Information, Hanover, MD, 2006), p. 82.

\bibitem{Sneden2003}
C. Sneden et al., Astrophys. J. {\bf 591}, 936 (2003).

\bibitem{Zhao2010}
P.~W. Zhao, Z.~P. Li, J.~M. Yao, and J. Meng, Phys. Rev. C {\bf 82}, 054319 (2010).

\bibitem{Meng2006}
J. Meng, H. Toki, S.~G. Zhou, S.~Q. Zhang, W.~H. Long, and L.~S. Geng, Prog. Part. Nucl. Phys. {\bf 57}, 470 (2006).

\bibitem{Hill2002}
V. Hill et al., Astron. Astrophys. {\bf 387}, 560 (2002).

\bibitem{Sneden2009}
C. Sneden, J.~E. Lawler, J.~J. Cowan, I.~I. Ivans, and E.~A. D. Hartog,
Astrophys. J. Suppl. Ser. {\bf 182}, 80 (2009).

\bibitem{Roederer2012a}
I.~U. Roederer, J.~E. Lawler, J.~J. Cowan, T.~C. Beers, A. Frebel, I.~I. Ivans,
H. Schatz, J.~S. Sobeck, and C. Sneden, Astrophys. J. {\bf 747}, L8 (2012).

\bibitem{Roederer2012b}
I.~U. Roederer and J.~E. Lawler, Astrophys. J. {\bf 750}, 76 (2012).

\bibitem{Qian2001}
Y.-Z. Qian and G.Wasserburg, Astrophys. J. {\bf 559}, 925 (2001).

\bibitem{Travaglio2004}
C. Travaglio, R. Gallino, E. Arnone, J. Cowan, F. Jordan, and C. Sneden, Astrophys. J. {\bf 601}, 864 (2004).

\bibitem{Ishimaru2005}
Y. Ishimaru, S. Wanajo, W. Aoki, S.~G. Ryan, and N. Prantzos, Nucl. Phys. A {\bf 758}, 603 (2005).

\bibitem{Montes2007}
F. Montes et al., Astrophys. J. {\bf 671}, 1685 (2007).

\bibitem{Arcones2011b}
A. Arcones and F. Montes, Astrophys. J. {\bf 731}, 5 (2011).

\bibitem{Honda2006}
S. Honda, W. Aoki, Y. Ishimaru, S. Wanajo, and S.~G. Ryan, Astrophys. J. {\bf 643}, 1180 (2006).

\bibitem{Honda2007}
S. Honda, W. Aoki, Y. Ishimaru, and S. Wanajo, Astrophys. J. {\bf 666}, 1189 (2007).

\bibitem{Fu2011}
G.~J. Fu, Y. Lei, H. Jiang, Y.~M. Zhao, B. Sun, and A. Arima, Phys. Rev. C {\bf 84}, 034311 (2011).

\bibitem{Sun2011}
B. Sun, P.~W. Zhao and J. Meng, Sci. China Ser. G: Phys., Mech. Astron. {\bf 54}, 210 (2011).

\bibitem{Mumpower2012a}
M.~R. Mumpower, G.~C. McLaughlin, and R. Surman, Astrophys. J. {\bf 752}, 117 (2012).

\bibitem{Mumpower2012b}
M.~R. Mumpower, G.~C. McLaughlin, and R. Surman, Phys. Rev. C {\bf 85}, 045801 (2012).
\end{thebibliography}

\end{CJK*}
\end{document}